\documentclass[11pt]{article}
\usepackage{geometry}
\usepackage{graphicx}
\usepackage[section]{placeins}
\usepackage{microtype}
\usepackage{amssymb}
\usepackage{amsmath}
\usepackage{epstopdf}
\usepackage{cite}
\usepackage{bbm}
\usepackage{multirow}
\usepackage{enumerate}
\usepackage{ulem}
\usepackage{xcolor}
\usepackage{appendix}
\usepackage[hyperindex=true,
          pdfstartview=FitH,
          bookmarksnumbered=true,
          bookmarksopen=true,
          citecolor=blue,
          linkcolor=blue,
          colorlinks=true,
          pdfborder={0 0 0}]{hyperref}

\newcommand{\ba}{\begin{align}}
\newcommand{\ea}{\end{align}}
\newcommand{\be}{\begin{equation}}
\newcommand{\en}{\end{equation}}
\newcommand{\bea}{\begin{eqnarray}}
\newcommand{\ena}{\end{eqnarray}}

\title{Turning-Point Count Discrepancy as a Diagnostic of Relativistic Orbital Chaos}

\author{
Wenfu Cao$^{1}$\footnote{202411100001@stu.ujn.edu.cn},
Ying Wang$^{2}$\footnote{wangying424524@163.com},
Hongsheng Zhang$^{1}$\footnote{sps\_zhanghs@ujn.edu.cn (corresponding author)}
\\[0.8em]
\small
$^{1}$School of Physics and Technology, University of Jinan,\\
\small
336 West Road of Nan Xinzhuang, Jinan, Shandong 250022, China
\\[0.4em]
\small
$^{2}$School of Mathematics, Physics and Statistics,\\
\small
Shanghai University of Engineering Science, Shanghai 201620, China
}

\begin{document}

\maketitle

\begin{abstract}
We propose the turning-point count-discrepancy indicator (TPCD) for diagnosing orbital chaos from a single trajectory in relativistic Hamiltonian systems with two oscillatory degrees of freedom. TPCD measures the largest cumulative departure of one turning-event count from its mean rate per reference cycle, requiring neither a neighboring orbit nor a phase-space partition and applying to both massive particles and photons. We establish its long-time behavior under explicit event--phase assumptions. Rigid phases with an exact event--phase correspondence obey a strict discrepancy bound of unity, and linearizable regular tori with bounded degree-one phase deformations obey a finite, orbit-dependent bound; both imply that the normalized indicator decays to zero as the record grows. A diffusive fluctuation mechanism instead yields a Brownian-bridge scaling and a finite statistical scale. Integrable Kerr motion validates the construction, recovering prescribed frequency ratios from event counts to within $2.6\times10^{-5}$ for six targets, including an irrational ratio. In charged-particle scans around a Kerr black hole in an external test magnetic field, TPCD and the fast Lyapunov indicator agree for all 80 sampled trajectories. In the Schwarzschild--Melvin photon model, a trajectory with elevated finite-time TPCD but low fast-Lyapunov values is identified as regular once its indicator trends downward over an extended integration, showing that finite-time values must be read together with their long-time trend.
\end{abstract}

\section{Introduction}\label{sec:introduction}
Relativistic orbital dynamics offers a setting in which integrable motion, resonances, and chaos can coexist within closely related physical models. Once stationarity and axial symmetry have been used to fix the energy and the axial angular momentum, many orbital problems reduce to Hamiltonian systems with two degrees of freedom: the radial and polar oscillations may be organized on invariant tori, or they may be strongly coupled, so that the accessible phase space is mixed and contains both regular islands and chaotic regions. Kerr geodesics provide the canonical integrable reference, because the Carter constant renders the Hamilton--Jacobi equation separable \cite{Carter:1968rr,Wilkins:1972rs}. Two physically distinct departures from this reference are of particular interest here. Charged-particle motion around a Kerr black hole in an asymptotically uniform magnetic field introduces electromagnetic forces on a fixed geometry \cite{Wald:1974np}, whereas photon motion in the Schwarzschild--Melvin spacetime incorporates the gravitational effect of the magnetic field into the geometry itself \cite{Ernst:1976,LimaJunior:2021,Li:2018wtz}.

Numerous studies have used Poincar\'e sections, Lyapunov exponents, and various fast chaos indicators to identify and characterize chaotic motion in these and other nonintegrable relativistic models, establishing a broad set of diagnostic tools for orbital chaos
\cite{Wu:2003pe,Wu:2006rx,Ma:2014aha,Cao:2024ihv,Liu:2026npn,Liu:2026mur,Wang:2026whj,Xu:2026oic,Lu:2026kcm,Li:2026ocp,Zhang:2025qia,Liu:2025izb,Xu:2024ble,Lu:2024srb,Das:2024iuf,Yi:2020shw,Chen:2016tmr}.
Recurrence and information-theoretic approaches provide complementary descriptions of orbital organization \cite{Cao:2024bjk,Cao:2024rvo,Cao:2026qvy,Kopacek:2010yr}. Each construction emphasizes a different feature: the geometry of a section, the growth of nearby-orbit separation, or the statistical structure of a trajectory record. Most of them draw on one of three resources that are not always available: a phase-space section, a second trajectory evolved alongside the first, or a record long enough for a statistical estimator to converge. A diagnostic that is intrinsic to a single trajectory and sensitive to the relative oscillatory structure of its two degrees of freedom would therefore be complementary to the existing toolkit.

Here we construct such a diagnostic. Selecting only one orientation of each turning event, for instance the local minima, gives two ordered event sequences. The events of the first coordinate define a reference-cycle counter, and the events of the second supply an integer count within each reference cycle. A stable relation between the two phases constrains the cumulative count to remain close to a line; the relevant departure is signed, records both an excess and a deficit of events, and its largest absolute value measures the departure over an entire observation interval. The same record also provides a mean event-count ratio, but the ratio and the cumulative discrepancy carry different information: a trajectory can remain close to a rational average ratio while exhibiting substantial accumulated changes in its event sequence. We therefore define the turning-point count-discrepancy indicator (TPCD) as the largest cumulative discrepancy normalized by the diffusive scale expected for the number of events observed.

The central theoretical statement of this paper is a bounded cumulative discrepancy under explicit event--phase assumptions. Exact correspondence with rigid phases gives the strict bound $|R_m^{(N)}|<1$, which is a counting bound rather than a chaos threshold. Bounded, degree-one deformations of the phases of a linearizable regular torus give a finite but orbit-dependent bound, so that a large finite-time discrepancy does not by itself contradict regular motion. Both results imply that the normalized indicator tends to zero for such regular trajectories as the record grows. On the chaotic side, a diffusive fluctuation mechanism provides a contrasting statistical expectation, in which TPCD retains a finite scale instead of decaying. We test these expectations in two steps. We first validate the event-count ratio using integrable Kerr motion, and we then examine charged-particle trajectories around a Kerr black hole in an external test magnetic field and Schwarzschild--Melvin photon trajectories. The Kerr scans show agreement between TPCD and the fast Lyapunov indicator, while the photon examples illustrate why trajectories with elevated finite-time TPCD may require longer observations and an independent diagnostic to establish their dynamical character.

The remainder of this paper is organized as follows. Section~\ref{sec:framework} develops the general event-count framework, derives the three asymptotic regimes, and defines TPCD. Section~\ref{sec:kerr} presents the integrable Kerr benchmark and the charged-particle application in an external test magnetic field. Section~\ref{sec:melvin} introduces Schwarzschild--Melvin photon dynamics and examines an energy scan together with extended integrations. Section~\ref{sec:conclusions} summarizes the results and their scope. We use geometrized units $G=c=M=1$; the unit particle rest mass applies only to the timelike Kerr examples.

\section{General event-count framework and TPCD}\label{sec:framework}

The construction requires two recurring oscillatory event sequences, rather than a particular metric. Let $\eta$ denote an increasing evolution parameter. It may be proper time for a massive particle or an affine parameter for a photon. A monotonic reparametrization preserves the event ordering and hence the discrepancy for a fixed event record, although it changes the parameter value at which that record is reached.

\subsection{Event sequences and cumulative counts}\label{sec:events}
For each of two oscillatory coordinates, select a single orientation of its simple turning points, such as a local minimum. Let the selected events of the first coordinate occur at
\begin{equation}
\eta_0<\eta_1<\cdots<\eta_N.
\label{eq:event_times}
\end{equation}
These events delimit $N$ complete reference cycles. We use the half-open intervals $[\eta_{i-1},\eta_i)$ and denote the number of selected second-coordinate events in the $i$th interval by $q_i$. An event exactly on a boundary belongs to the newly started interval. Initial and final incomplete reference cycles are excluded. The cumulative count and its sample mean are
\begin{equation}
C_0=0,\qquad C_m=\sum_{i=1}^m q_i,\qquad
\overline q_N=\frac{C_N}{N}.
\label{eq:counts}
\end{equation}
The signed cumulative discrepancy is
\begin{equation}
R_m^{(N)}=C_m-\frac{m}{N}C_N
=\sum_{i=1}^m(q_i-\overline q_N),\qquad 0\leq m\leq N,
\label{eq:residual}
\end{equation}
and its largest magnitude is
\begin{equation}
R_{\max}(N)=\max_{0\leq m\leq N}|R_m^{(N)}|.
\label{eq:max_discrepancy}
\end{equation}
The superscript emphasizes that extending the record changes the sample mean and therefore recenters all preceding residuals. In particular, $R_0^{(N)}=R_N^{(N)}=0$, so the terminal residual itself carries no information about intermediate departures. The absolute value treats excess and deficit counts on equal terms. It is taken after summation and therefore preserves cancellations within the cumulative discrepancy.

If $C_m/m\to\beta>0$, then $\beta$ is the mean number of second-coordinate events per reference cycle. Under a one-to-one correspondence between events and phase cycles, the reciprocal ratio
\begin{equation}
\rho=\frac{\omega_1}{\omega_2}=\frac1\beta,
\qquad \widehat\rho_N=\frac{N}{C_N}
\label{eq:ratio_convention}
\end{equation}
is estimated by the cycle counts. In the applications, coordinate 1 is radial and coordinate 2 is polar, so $N=N_r$, $C_N=N_\theta$, and the plotted ratio is always $N_r/N_\theta$. Without the event--phase correspondence, this quantity remains an event-count ratio and need not equal a fundamental frequency ratio. A $p{:}q$ radial-to-polar resonance denotes $\rho=p/q$; a nearby finite-time event-count ratio alone does not establish a resonance.
In both applications, we select turning events at which $p_r$ or $p_\theta$ changes from negative to positive. Outside the horizon, these events correspond to local minima of $r$ and $\theta$, respectively. Two successive selected radial events delimit one complete radial cycle, within which the selected polar events are counted.

\subsection{Rigid phases and the strict discrepancy bound}\label{sec:strict}
Suppose that an integrable trajectory admits phases that advance linearly in an evolution parameter,
\begin{equation}
\psi_i(\eta)=\psi_i(\eta_0)+\omega_i(\eta-\eta_0),
\qquad \omega_i>0,\quad i=1,2.
\label{eq:rigid_phases}
\end{equation}
We additionally require that the selected turning events occur exactly when the corresponding phase crosses $2\pi\mathbb Z$, once per complete phase cycle. This is a condition on the physical events; the existence of action--angle coordinates alone does not ensure it for arbitrary coordinate extrema.

At the end of $m$ reference cycles, the second phase has advanced by $2\pi\beta m$, where $\beta=\omega_2/\omega_1$. The half-open convention therefore gives the exact integer count
\begin{equation}
C_m=\left\lceil\chi+\beta m\right\rceil-\lceil\chi\rceil,
\qquad \chi=\frac{\psi_2(\eta_0)}{2\pi}.
\label{eq:rigid_count}
\end{equation}
This expression also covers coincident events. Defining the rounding error
\begin{equation}
e(x)=\lceil x\rceil-x,\qquad 0\leq e(x)<1,
\label{eq:rounding_error}
\end{equation}
we obtain
\begin{equation}
C_m=\beta m+d_m,\qquad
d_m=e(\chi+\beta m)-e(\chi).
\label{eq:rigid_count_decomposition}
\end{equation}
Substitution into Eq.~\eqref{eq:residual} yields
\begin{equation}
R_m^{(N)}=e(\chi+\beta m)
-\left[\left(1-\frac mN\right)e(\chi)
+\frac mN e(\chi+\beta N)\right].
\label{eq:strict_residual}
\end{equation}
Both the first term and the bracketed convex combination belong to $[0,1)$. Consequently,
\begin{equation}
|R_m^{(N)}|<1,\qquad R_{\max}(N)<1.
\label{eq:strict_bound}
\end{equation}
The constant 1 is a strict counting-discrepancy bound under the stated assumptions, not a chaos threshold. The argument applies to rational and irrational $\beta$. It also implies $q_i\in\{\lfloor\beta\rfloor,\lceil\beta\rceil\}$. The assumptions exclude degenerate oscillations without recurring simple turning events.

\subsection{Regular tori in a nonintegrable system}\label{sec:regular_tori}
Consider a regular invariant torus of a possibly nonintegrable Hamiltonian system. Assume that its flow is conjugate to a linear flow, with lifted torus angles
\begin{equation}
\boldsymbol\vartheta(\eta)=\boldsymbol\vartheta(\eta_0)
+\boldsymbol\nu(\eta-\eta_0),\qquad \nu_1,\nu_2>0.
\label{eq:torus_flow}
\end{equation}
Assume further that the selected events are represented by lifted phases
\begin{equation}
\Phi_i(\boldsymbol\vartheta)=\vartheta_i+g_i(\boldsymbol\vartheta),
\qquad i=1,2,
\label{eq:deformed_phases}
\end{equation}
where each $g_i$ is continuous and $2\pi$-periodic in both angles. Each crossing of the next integer phase level corresponds to exactly one selected event, with no extra crossings or missing events. This degree-one event--phase correspondence is a sufficient hypothesis, not an automatic consequence of regularity.

Let $\beta=\nu_2/\nu_1$ and $G=g_2-\beta g_1$. The linear terms cancel to give
\begin{equation}
\begin{split}
&\Phi_2(\eta)-\Phi_2(\eta_0)
-\beta[\Phi_1(\eta)-\Phi_1(\eta_0)]\\
&\hspace{1cm}=G(\boldsymbol\vartheta(\eta))-G(\boldsymbol\vartheta(\eta_0)).
\end{split}
\label{eq:phase_difference}
\end{equation}
Continuity on the compact torus makes $G$ bounded. Since the first phase advances by $2\pi m$ at $\eta_m$, the second phase advances by $2\pi\beta m$ plus a bounded correction. Conversion to the integer event count adds a rounding error of magnitude less than 1. Thus
\begin{equation}
C_m=\beta m+\varepsilon_m,\qquad
|\varepsilon_m|\leq K,
\label{eq:bounded_count_error}
\end{equation}
where a finite choice is $K=\|G\|_\infty/\pi+1$, independent of $m$. It follows that
\begin{equation}
\overline q_N=\beta+\frac{\varepsilon_N}{N},\qquad
R_m^{(N)}=\varepsilon_m-\frac mN\varepsilon_N,
\label{eq:torus_residual}
\end{equation}
and hence
\begin{equation}
|R_m^{(N)}|\leq |\varepsilon_m|+\frac mN|\varepsilon_N|
\leq2K,\qquad R_{\max}(N)\leq2K.
\label{eq:torus_bound}
\end{equation}
The constant $K$ can differ between tori and can be large. A large finite-time discrepancy therefore does not by itself contradict regular motion. When coordinate extrema are created or lost, or the selected events have a different winding structure, the assumed correspondence must be reconsidered. Neither Eq.~\eqref{eq:strict_bound} nor Eq.~\eqref{eq:torus_bound} is asserted for every regular trajectory without these hypotheses.

\subsection{Chaotic count fluctuations: a statistical mechanism}\label{sec:diffusion}
On a chaotic trajectory, the number of second-coordinate events per reference cycle may change irregularly. To describe a possible statistical regime, suppose that $q_i$ has a stationary mean $\mu>0$ and define $X_i=q_i-\mu$ and $S_m=\sum_{i=1}^mX_i$. Sample centering gives the exact relation
\begin{equation}
R_m^{(N)}=S_m-\frac mN S_N.
\label{eq:centered_partial_sum}
\end{equation}
Finite variance and sufficiently weak correlations motivate a diffusive model. A precise sufficient statistical assumption is a functional central limit law,
\begin{equation}
\frac{S_{\lfloor Nu\rfloor}}{\sigma_q\sqrt N}
\ \Longrightarrow\ W(u),\qquad 0\leq u\leq1,
\label{eq:functional_limit}
\end{equation}
with positive effective variance $\sigma_q^2$ and standard Brownian motion $W$. Under this assumption, subtracting the fitted mean produces a Brownian bridge rather than an unconditioned random walk:
\begin{equation}
\frac{R_{\lfloor Nu\rfloor}^{(N)}}{\sigma_q\sqrt N}
\ \Longrightarrow\ W(u)-uW(1).
\label{eq:bridge_limit}
\end{equation}
The typical maximum discrepancy then scales as $\sqrt N$. This is a conditional statistical mechanism for an event observable, not a theorem that every chaotic trajectory has unbounded count discrepancy. Long correlations, sticky motion, coherent drift, or an event sequence weakly coupled to the chaotic degrees of freedom can give different behavior.

\subsection{TPCD and observation-time dependence}\label{sec:tpcd_definition}
For $C_N>0$, define
\begin{equation}
I_{\mathrm{TPCD}}(N)=\frac{R_{\max}(N)}{\sqrt{C_N}}
=\frac{\displaystyle\max_{0\leq m\leq N}
\left|\sum_{i=1}^m(q_i-\overline q_N)\right|}{\sqrt{C_N}}.
\label{eq:tpcd}
\end{equation}
Event counts are already dimensionless. The role of $\sqrt{C_N}$ is to normalize by the expected diffusive scale, not to confer dimensional consistency. For the rigid-phase and regular-torus cases, respectively,
\begin{equation}
I_{\mathrm{TPCD}}(N)<\frac1{\sqrt{C_N}},\qquad
I_{\mathrm{TPCD}}(N)\leq\frac{2K}{\sqrt{C_N}}
\longrightarrow0,
\label{eq:regular_tpcd_limit}
\end{equation}
as $C_N\to\infty$. In the statistical regime of Sec.~\ref{sec:diffusion}, if $C_N/N\to\mu$, then
\begin{equation}
I_{\mathrm{TPCD}}(N)\ \Longrightarrow\
\frac{\sigma_q}{\sqrt\mu}\sup_{0\leq u\leq1}|W(u)-uW(1)|.
\label{eq:diffusive_tpcd_limit}
\end{equation}
Here the arrow denotes convergence in distribution. The indicator retains a nonzero statistical scale, rather than converging to a prescribed constant along every trajectory.

The practical distinction suggested by Eqs.~\eqref{eq:regular_tpcd_limit} and \eqref{eq:diffusive_tpcd_limit} is therefore based on the long-time behavior of TPCD. For a regular trajectory satisfying the assumptions of Secs.~\ref{sec:strict} or \ref{sec:regular_tori}, the normalized discrepancy decreases toward zero as the observation record grows. By contrast, when the event-count fluctuations are diffusive, TPCD remains on a finite statistical scale and continues to fluctuate rather than showing sustained decay. Thus, a trajectory whose TPCD decreases toward zero is identified as a regular candidate, whereas a trajectory whose TPCD remains persistently finite over increasingly long records is identified as a chaotic candidate. Because finite-time regular trajectories with a large bound $K$, as well as sticky chaotic trajectories, may show intermediate behavior, this distinction should be checked over increasing observation times and, when possible, against an independent chaos diagnostic.

In a specific parameter scan, clear separation of TPCD values, together with agreement with an independent chaos diagnostic, can support the identification of the sampled trajectories as regular or chaotic. Trajectories for which this separation is unclear or the diagnostics disagree require further examination over longer observation intervals.

\section{Kerr benchmark and charged-particle motion in an external magnetic field}\label{sec:kerr}

The integrable Kerr problem provides an explicit realization of the rigid-phase assumptions in Sec.~\ref{sec:strict}. We use it first to connect event counts to independently computed periods. The same geometric background then supports a nonintegrable charged-particle application when an external test magnetic field is introduced.

\subsection{Integrable Kerr motion and frequency-ratio validation}\label{sec:kerr_benchmark}
In Boyer--Lindquist coordinates $(t,r,\theta,\phi)$, the Kerr metric is
\begin{equation}
\begin{split}
ds^2={}&-\left(1-\frac{2r}{\Sigma}\right)dt^2
-\frac{4ar\sin^2\theta}{\Sigma}\,dt\,d\phi
+\frac{\Sigma}{\Delta}\,dr^2+\Sigma\,d\theta^2\\
&+\left(r^2+a^2+\frac{2a^2r\sin^2\theta}{\Sigma}\right)
\sin^2\theta\,d\phi^2,
\label{eq:kerr_metric}
\end{split}
\end{equation}
where $\Sigma=r^2+a^2\cos^2\theta$ and $\Delta=r^2-2r+a^2$. For a timelike geodesic of unit rest mass, the conserved quantities are $E=-p_t$, $L=p_\phi$, and the Carter constant $Q$ \cite{Carter:1968rr}. With proper time $\tau$, the separated equations are
\begin{equation}
\Sigma^2\left(\frac{dr}{d\tau}\right)^2=\mathcal R(r),\qquad
\Sigma^2\left(\frac{d\theta}{d\tau}\right)^2=\Theta(\theta),
\label{eq:kerr_separated_motion}
\end{equation}
where
\begin{align}
\mathcal R(r)&=[E(r^2+a^2)-aL]^2
-\Delta[r^2+(L-aE)^2+Q],\label{eq:radial_potential}\\
\Theta(\theta)&=Q-\cos^2\theta\left[a^2(1-E^2)+\frac{L^2}{\sin^2\theta}\right].
\label{eq:polar_potential}
\end{align}
The radial potential is written as $\mathcal R$ to distinguish it from the count residual $R_m^{(N)}$. The associated actions can be written as
\begin{equation}
J_r=\frac1\pi\int_{r_{\rm p}}^{r_{\rm a}}\frac{\sqrt{\mathcal R(r)}}{\Delta}\,dr,
\qquad
J_\theta=\frac1\pi\int_{\theta_{\min}}^{\theta_{\max}}\sqrt{\Theta(\theta)}\,d\theta,
\label{eq:kerr_actions}
\end{equation}
where the integration limits are adjacent turning points \cite{Hinderer:2008dm,Kerachian:2023oiw}.

For establishing the event correspondence explicitly, use Mino time $\lambda$, defined by $d\tau=\Sigma\,d\lambda$ \cite{Mino:2003yg,Drasco:2003ky}. Then $(dr/d\lambda)^2=\mathcal R$ and $(d\theta/d\lambda)^2=\Theta$. The complete radial and polar periods are
\begin{equation}
\Lambda_r=2\int_{r_{\rm p}}^{r_{\rm a}}\frac{dr}{\sqrt{\mathcal R(r)}},\qquad
\Lambda_\theta=2\int_{\theta_{\min}}^{\theta_{\max}}\frac{d\theta}{\sqrt{\Theta(\theta)}}.
\label{eq:mino_periods}
\end{equation}
The factor of 2 includes the outgoing and returning parts of an oscillation. The frequencies and phases are
\begin{equation}
\Upsilon_i=\frac{2\pi}{\Lambda_i},\qquad
w_i(\lambda)=w_i(\lambda_0)+\Upsilon_i(\lambda-\lambda_0),
\qquad i=r,\theta.
\label{eq:mino_frequencies}
\end{equation}
Choose the phase origins so that $w_r=2\pi n$ marks $p_r:-\to+$ and $w_\theta=2\pi k$ marks $p_\theta:-\to+$. Each selected event then corresponds exactly to one complete cycle. These separated phases need not be identified with canonical proper-time angles. Since $\Sigma>0$, changing between proper time and Mino time preserves the geometric orbit and event ordering. It follows that
\begin{equation}
\rho_0=\frac{\Upsilon_r}{\Upsilon_\theta}
=\frac{\Lambda_\theta}{\Lambda_r}
=\lim_{N\to\infty}\frac{N}{C_N}.
\label{eq:frequency_validation_period_ratio}
\end{equation}
Both mean proper-time event frequencies acquire the same conversion factor, so their ratio is also $\rho_0$. No change of initial coordinates or momenta is needed when the same orbit is integrated in proper time.

For the frequency-ratio validation, we set $b=0$ and fix $E=0.98$, $L=2$, and $a=0.99$. Each trajectory starts at a radial periastron, $r_0=r_{\rm p}$, with $\theta_0=\pi/2$, $\phi_0=0$, and $p_{r0}=0$. The condition $\mathcal R(r_0)=0$ in Eq.~\eqref{eq:radial_potential} determines the Carter constant,
\begin{equation}
Q(r_0)=\frac{\left[E(r_0^2+a^2)-aL\right]^2}{\Delta(r_0)}
-r_0^2-(L-aE)^2,
\qquad p_{\theta0}=\sqrt{Q(r_0)}>0.
\label{eq:frequency_validation_initial_constraint}
\end{equation}
Using the radial and polar potentials with this $Q(r_0)$, we evaluate the Mino-time periods in Eq.~\eqref{eq:mino_periods}. The target is imposed through
Eq.~\eqref{eq:frequency_validation_period_ratio}.
We numerically solve for $r_0$ such that this ratio equals each prescribed target, selecting nondegenerate bound orbits with $r_0$ as the periastron and a finite outer apastron. Thus $r_0$ is the only independently adjusted initial coordinate; $Q$ and $p_{\theta0}$ follow from the constraint. Five targets are rational ratios, $1/2$, $2/3$, $3/4$, $4/5$, and $5/6$, and the sixth is the irrational target $1/\sqrt{2}$ for a quasiperiodic reference orbit. Since the selected event counts are unchanged by the monotonic transformation between Mino and proper time, they can be obtained directly from the proper-time numerical integration. The equations of motion are integrated using an eighth-order Runge--Kutta scheme with a fixed proper-time step $h=0.1$. The numerical accuracy is monitored through conservation of the Hamiltonian, whose deviation remains below $10^{-12}$ throughout the integrations.

\begin{table}[htbp]
    \centering
    \renewcommand{\arraystretch}{1.2}
    \setlength{\tabcolsep}{12pt}
    \caption{Validation of the radial-to-polar frequency ratio using integrable Kerr trajectories. Initial radii are obtained by solving Eq.~\eqref{eq:frequency_validation_period_ratio} at $E=0.98$, $L=2$, and $a=0.99$. The last column gives the event-count ratios from the Fortran integrations.}
    \label{tab:frequency_ratio_validation}
    \begin{tabular}{crr}
        \hline
        Theoretical target ratio & Initial radius $r_0$ & Computed $N_r/N_\theta$ \\
        \hline
        $1/2$        & 2.087066832629450 & 0.500000000000000 \\
        $2/3$        & 3.626483251540520 & 0.666666666666667 \\
        $3/4$        & 5.394765043695204 & 0.749984989492645 \\
        $4/5$        & 7.377636403973950 & 0.800025604916144 \\
        $5/6$        & 9.641851772978375 & 0.833355552592988 \\
        $1/\sqrt{2}$ & 4.334774218362792 & 0.707101597009854 \\
        \hline
    \end{tabular}
\end{table}

Table~\ref{tab:frequency_ratio_validation} shows close agreement between the prescribed theoretical ratios and the measured $N_r/N_\theta$, with absolute differences below $2.6\times10^{-5}$ for all six trajectories. The small differences are consistent with finite cycle-count resolution and numerical integration errors. The irrational-target example also confirms that the counting method is not restricted to commensurate radial--polar motion.

\subsection{Charged particles in an external test magnetic field}\label{sec:kerr_field}

We introduce an asymptotically uniform magnetic field aligned with the rotation axis of the black hole. In the test-field approximation, the electromagnetic field affects the charged-particle motion but does not backreact on the Kerr geometry. For an uncharged black hole, the corresponding Wald vector potential is \cite{Wald:1974np}
\begin{equation}
A_{\mu}
=
\frac{B}{2}
\left(g_{\mu\phi}+2a g_{\mu t}\right),
\label{eq:wald_potential}
\end{equation}
where $B$ is the magnetic-field strength measured at spatial infinity. The interaction between the particle and the magnetic field is characterized by the dimensionless coupling parameter
\begin{equation}
b=\frac{qB}{m},
\label{eq:magnetic_coupling}
\end{equation}
where $q$ is the particle charge. In the numerical calculations below, the charge-to-mass ratio is absorbed into the effective magnetic parameter $b$.

Let $P_\mu$ denote the canonical four-momentum. The charged-particle dynamics is governed by the super-Hamiltonian
\begin{equation}
\mathcal{H}
=
\frac{1}{2}g^{\mu\nu}
\left(P_{\mu}-qA_{\mu}\right)
\left(P_{\nu}-qA_{\nu}\right),
\label{eq:particle_hamiltonian}
\end{equation}
subject to the timelike mass-shell condition
\begin{equation}
\mathcal{H}=-\frac{m^{2}}{2}=-\frac{1}{2}.
\label{eq:mass_shell}
\end{equation}
Since neither the metric nor the vector potential depends explicitly on $t$ or $\phi$, the corresponding canonical momenta are conserved. The conserved energy and axial angular momentum are therefore
\begin{equation}
E=-P_{t},
\qquad
L=P_{\phi}.
\label{eq:energy_angular_momentum}
\end{equation}
Hereafter, $p_r$ and $p_\theta$ denote the radial and polar canonical momenta; they coincide with the corresponding mechanical momenta because $A_r=A_\theta=0$. After fixing $E$ and $L$, the motion reduces to a two-degree-of-freedom Hamiltonian system described by $(r,p_r,\theta,p_\theta)$. Its reduced Hamiltonian can be written as
\begin{equation}
\mathcal H_{\rm red}
=\frac{\Delta}{2\Sigma}p_r^2
+\frac{1}{2\Sigma}p_\theta^2
+V_{\rm eff}(r,\theta),
\label{eq:reduced_hamiltonian}
\end{equation}
where
\begin{equation}
\begin{split}
V_{\rm eff}(r,\theta)=\frac12\big[&
g^{tt}(E+qA_t)^2
-2g^{t\phi}(E+qA_t)(L-qA_\phi)\\
&+g^{\phi\phi}(L-qA_\phi)^2
\big].
\end{split}
\label{eq:effective_potential}
\end{equation}
The radial and polar coordinate velocities are
\begin{equation}
\dot r=\frac{\partial\mathcal H_{\rm red}}{\partial p_r}
=\frac{\Delta}{\Sigma}p_r,
\qquad
\dot\theta=\frac{\partial\mathcal H_{\rm red}}{\partial p_\theta}
=\frac{p_\theta}{\Sigma},
\label{eq:coordinate_velocities}
\end{equation}
where a dot denotes differentiation with respect to proper time $\tau$. Outside the event horizon, $\Delta>0$ and $\Sigma>0$; hence $p_r=0$ and $p_\theta=0$ identify radial and polar turning points, respectively.

The initial polar momentum is taken on the positive branch of the timelike constraint,
\begin{equation}
p_{\theta0}=\sqrt{-\Sigma_0-\Delta_0p_{r0}^2
-2\Sigma_0V_{\rm eff}(r_0,\theta_0)}.
\label{eq:kerr_initial_polar_momentum}
\end{equation}
The Kerr geometry remains fixed throughout this application. For the Wald potential in Eq.~\eqref{eq:wald_potential}, expansion at fixed $E,L$ gives
\begin{equation}
\mathcal H=\mathcal H_0+b\left(aE-\frac L2\right)
+\frac{b^2}{8}\left(g_{\phi\phi}+4ag_{t\phi}+4a^2g_{tt}\right).
\label{eq:wald_expansion}
\end{equation}
The term linear in $b$ is constant in the reduced problem; the quadratic term generally couples the radial and polar coordinates and destroys separability. The unperturbed Kerr action--angle transformation, where defined, can still be used as a coordinate transformation, but it does not make the perturbed actions constants of motion or supply global action--angle variables for the nonintegrable system. Regular trajectories of the perturbed system are covered by Sec.~\ref{sec:regular_tori} only when its torus and event hypotheses hold.

\subsection{Finite-time Kerr scans and independent comparisons}\label{sec:kerr_scans}
Both scans use $E=0.905$, $L=2$, $a=0.99$, $\theta_0=\pi/2$, and $p_{r0}=0$, with the positive $p_{\theta0}$ determined by the mass-shell constraint. An eighth-order Runge--Kutta method integrates the proper-time equations with step size $0.1$. The event-count ratio and TPCD are evaluated over an observation interval $T=10^7$, whereas the fast Lyapunov indicator (FLI) uses $T=10^6$. Cycles are counted through successive turning events at which the corresponding momentum changes from negative to positive, and only complete radial cycles enter TPCD. The plotted event-count ratio $N_r/N_\theta$, with the radial count in the numerator, estimates the consistently defined ratio $\rho=\omega_r/\omega_\theta$ when the stated event--phase correspondence holds. The radial-to-polar ratio and resonance convention are those of Eq.~\eqref{eq:ratio_convention}.

In both Kerr scans, TPCD clearly distinguishes regular and chaotic trajectories, in agreement with FLI. For the initial-radius scan, the largest TPCD value for regular trajectories is $0.016251$, whereas the smallest for chaotic trajectories is $4.376980$. For the magnetic-coupling scan, the corresponding values are $0.001776$ and $0.243879$. The wide separation between the two groups makes an imposed numerical threshold unnecessary for these data. The agreement with FLI across all 80 sampled trajectories, together with the representative Poincar\'e sections discussed below, demonstrates the effectiveness of TPCD in distinguishing regular and chaotic motion in these scans.

\begin{table}[htbp]
    \centering
    \renewcommand{\arraystretch}{1.15}
    \setlength{\tabcolsep}{10pt}
    \caption{Comparison of the identification of regular and chaotic trajectories by TPCD and FLI in the two Kerr scans. Both indicators give consistent results at all 80 sampled points: 57 regular trajectories and 23 chaotic trajectories.}
    \label{tab:classification_agreement}
    \begin{tabular}{lrrrr}
        \hline
        Scan & Sampled points & Both regular & Both chaotic & Disagreements \\
        \hline
        Initial radius $r_0$ & 40 & 26 & 14 & 0 \\
        Magnetic coupling $b$ & 40 & 31 & 9 & 0 \\
        Total & 80 & 57 & 23 & 0 \\
        \hline
    \end{tabular}
\end{table}

\begin{figure}[p]
    \centering
    \includegraphics[width=0.49\linewidth]{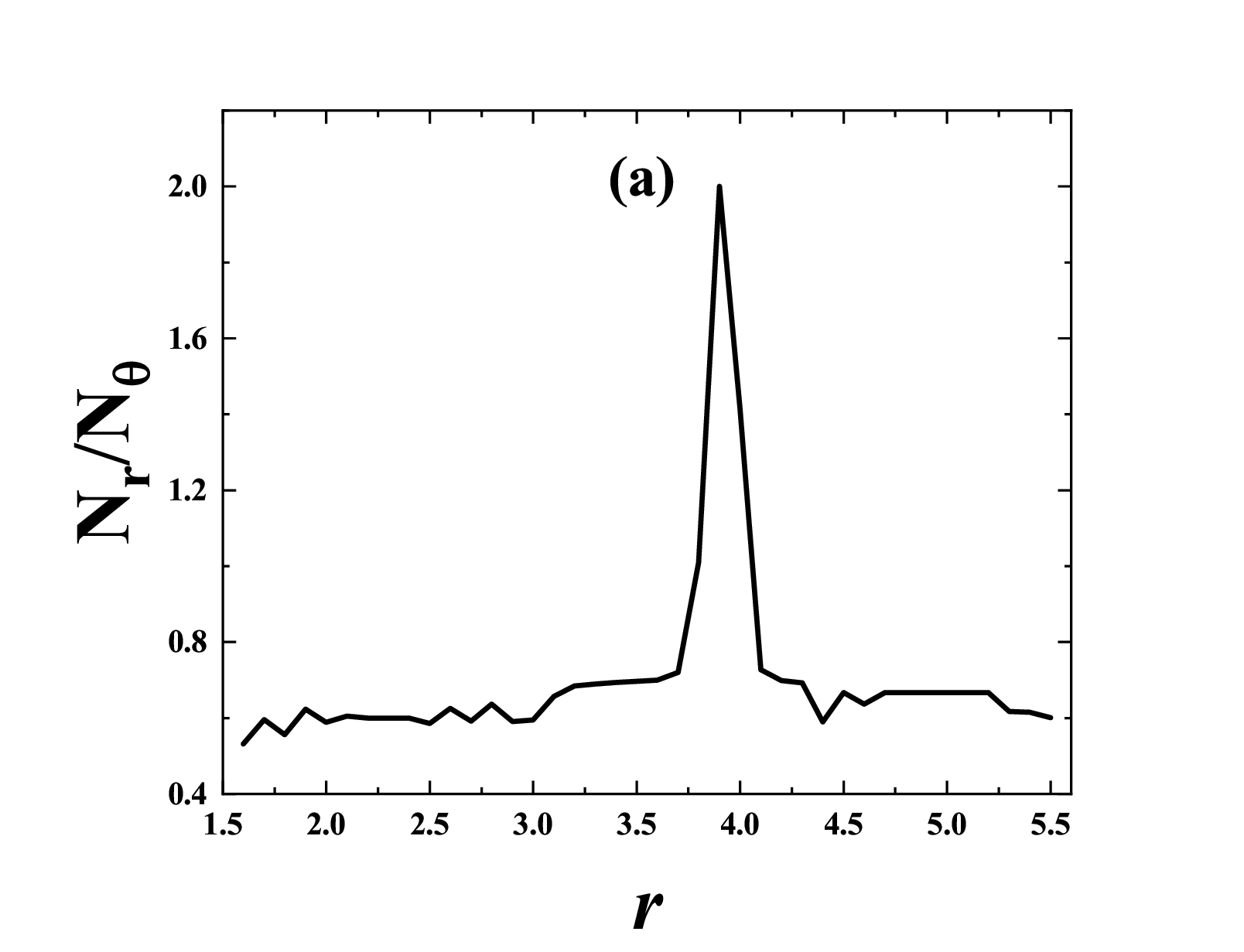}\hfill
    \includegraphics[width=0.49\linewidth]{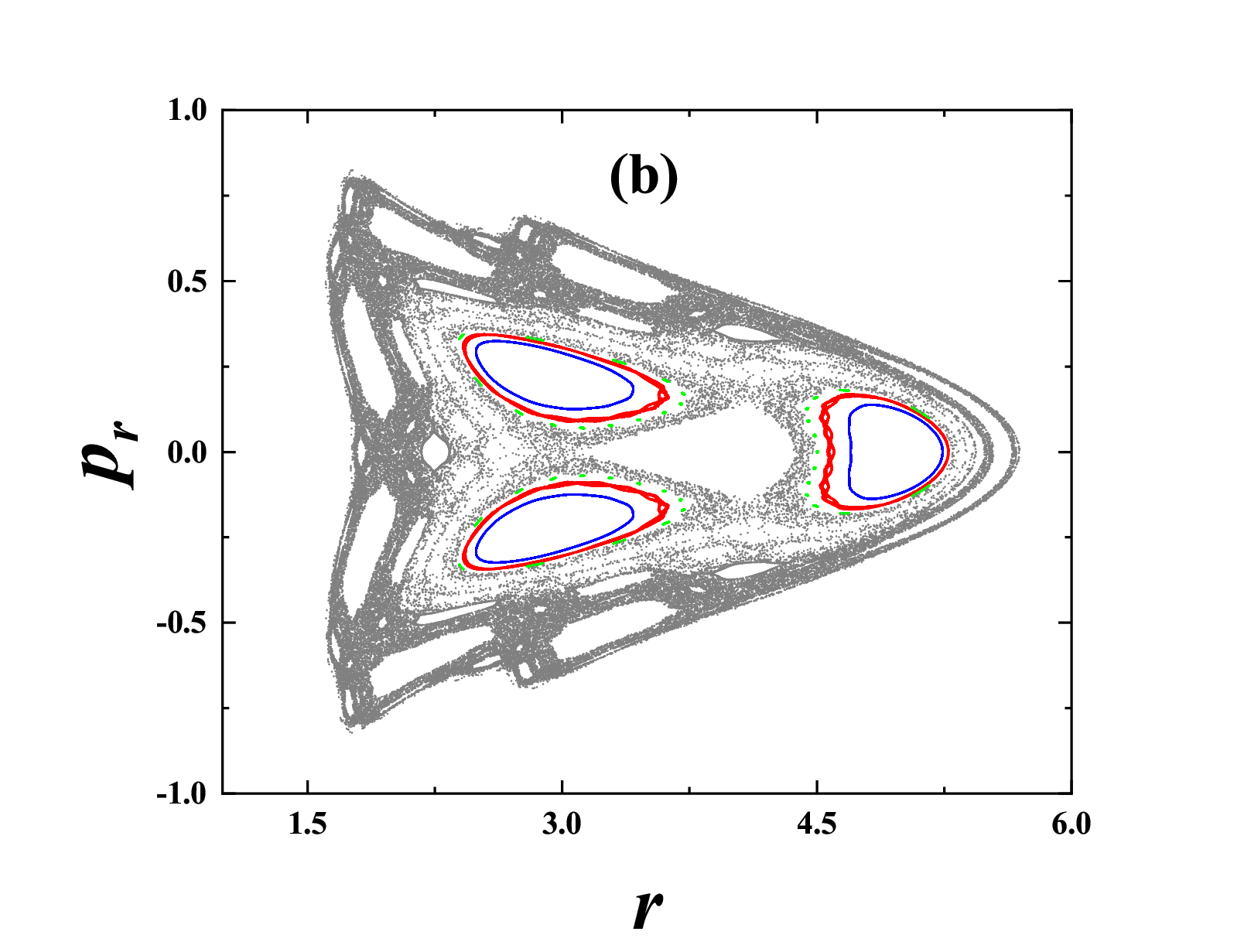}\\[2mm]
    \includegraphics[width=0.49\linewidth]{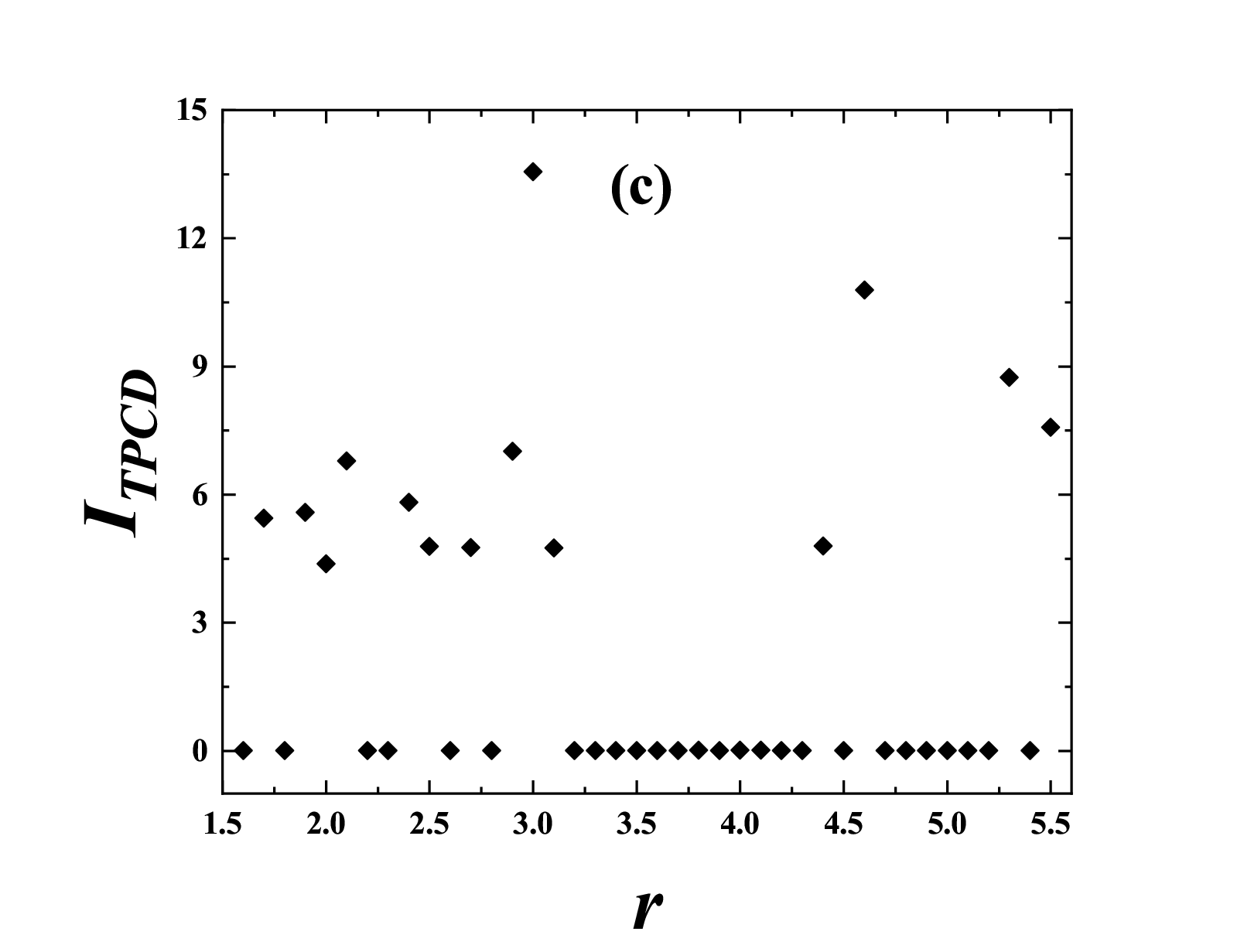}\hfill
    \includegraphics[width=0.49\linewidth]{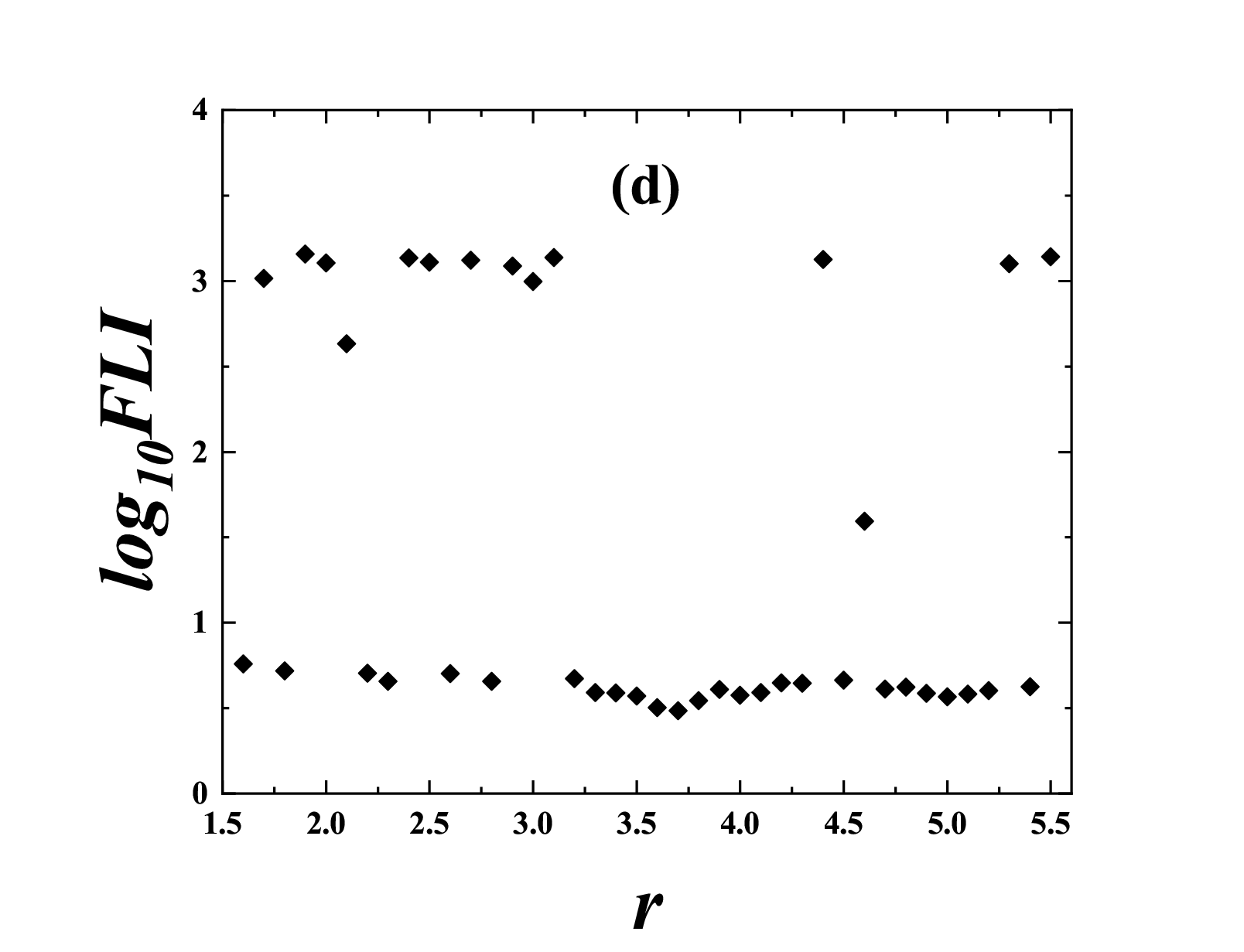}
    \caption{Initial-radius scan and representative Poincar\'e sections for charged-particle motion around a Kerr black hole immersed in an external magnetic field. The fixed parameters are $E=0.905$, $L=2$, $a=0.99$, and $b=0.105$. Initial conditions satisfy $\theta_0=\pi/2$, $p_{r0}=0$, and $p_{\theta0}>0$, with $p_{\theta0}$ determined by the mass-shell constraint. In panels (a), (c), and (d), the initial radius $r_0$ ranges from $1.6$ to $5.5$ in increments of $0.1$; the horizontal axis labeled $r$ denotes $r_0$. (a) Radial-to-polar event-count ratio $N_r/N_\theta$, obtained from same-direction turning events. (b) Representative Poincar\'e sections in the $(r,p_r)$ plane, sampled at $\theta=\pi/2$ with $\dot\theta>0$, showing a three-island structure and the surrounding scattered points. The gray, green, red, and blue points correspond to $r_0=4.4$, $4.5$, $4.6$, and $4.7$, respectively. (c) Turning-point count-discrepancy indicator $I_{\mathrm{TPCD}}$. (d) $\log_{10}(\mathrm{FLI})$ for the same initial-radius scan. The event-count ratio in panel (a) approximates $\omega_r/\omega_\theta$ when each selected turning-event sequence corresponds to complete cycles of the associated phase.}
    \label{fig:radius_scan_tpcd}
\end{figure}

Figure~\ref{fig:radius_scan_tpcd} scans 40 initial radii from $r_0=1.6$ to $5.5$ in increments of $0.1$ at fixed $b=0.105$. Panel (a) displays plateaus, abrupt changes, and a local peak. These changes in the event structure do not establish chaos: near $r_0=3.9$, the event-count ratio rises to approximately $2$, while both TPCD and FLI remain small.

Panels (c) and (d) of Fig.~\ref{fig:radius_scan_tpcd} show that TPCD and FLI consistently distinguish regular and chaotic trajectories in the initial-radius scan. Both indicators identify the same 14 chaotic trajectories, at $r_0=1.7$, $1.9$, $2.0$, $2.1$, $2.4$, $2.5$, $2.7$, $2.9$, $3.0$, $3.1$, $4.4$, $4.6$, $5.3$, and $5.5$, while the remaining 26 trajectories exhibit regular motion. The magnitudes of TPCD and FLI need not follow the same ordering because they measure different dynamical features: FLI measures the growth of separation between neighboring trajectories, whereas TPCD measures the cumulative discrepancy in the radial--polar event-count relation.

The Poincar\'e sections in Fig.~\ref{fig:radius_scan_tpcd}(b) further illustrate the orbital structure in the region where the event-count ratio is close to $2/3$. They are sampled at $\theta=\pi/2$ with $\dot\theta>0$ for $r_0=4.4$, $4.5$, $4.6$, and $4.7$. The green and blue trajectories, at $r_0=4.5$ and $4.7$, trace clear island curves, consistent with their small TPCD and FLI values. Their event-count ratios are close to $2/3$, consistent with a $2{:}3$ radial-to-polar resonance structure. The gray trajectory at $r_0=4.4$ produces section points distributed over a wider region and has large values of both indicators. Although the red trajectory at $r_0=4.6$ remains close to the island structure in the displayed section, both TPCD and FLI indicate its chaotic character. Thus, proximity to a resonance island or an event-count ratio close to a simple rational number does not guarantee regular motion.

\begin{figure}[p]
    \centering
    \includegraphics[width=0.49\linewidth]{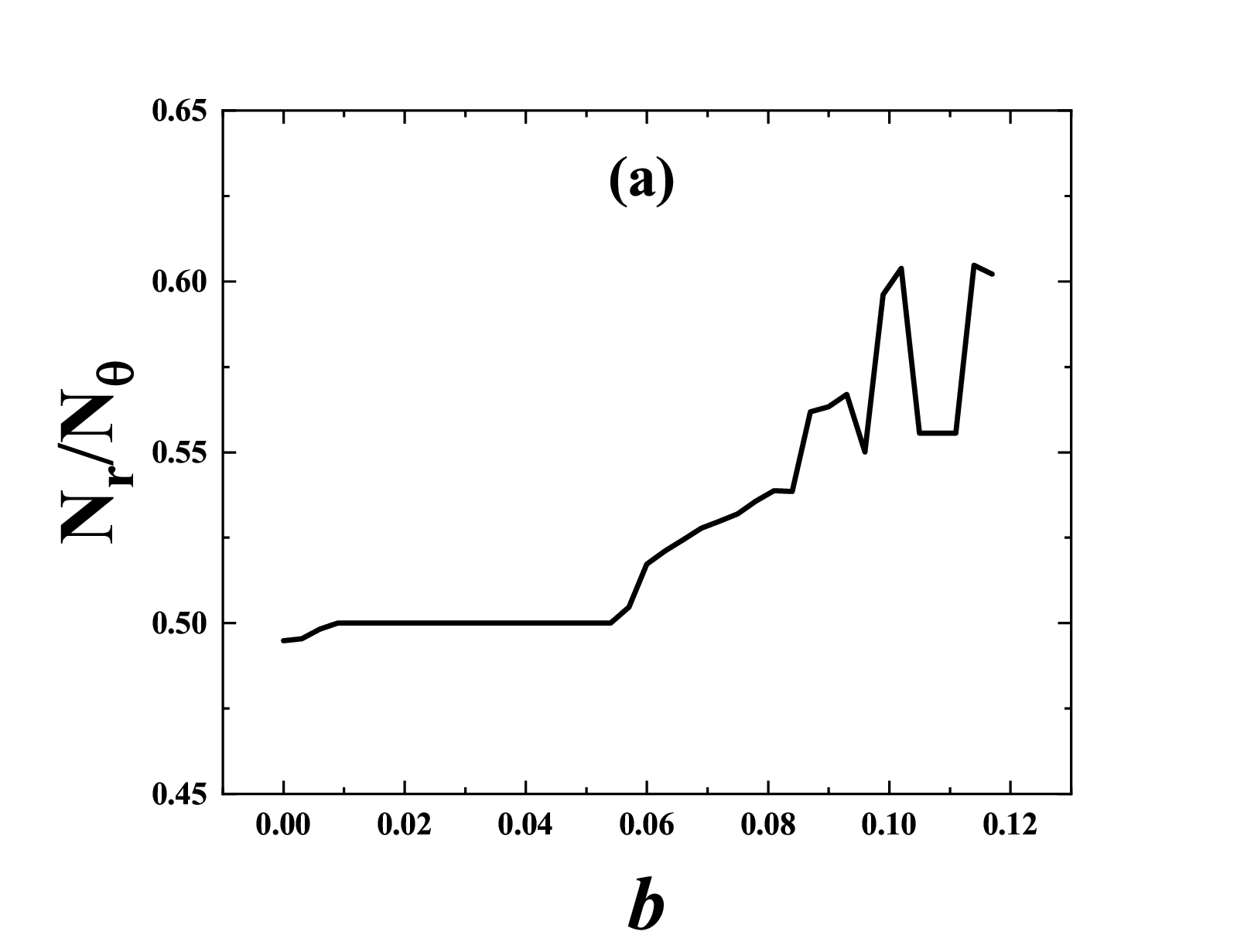}\hfill
    \includegraphics[width=0.49\linewidth]{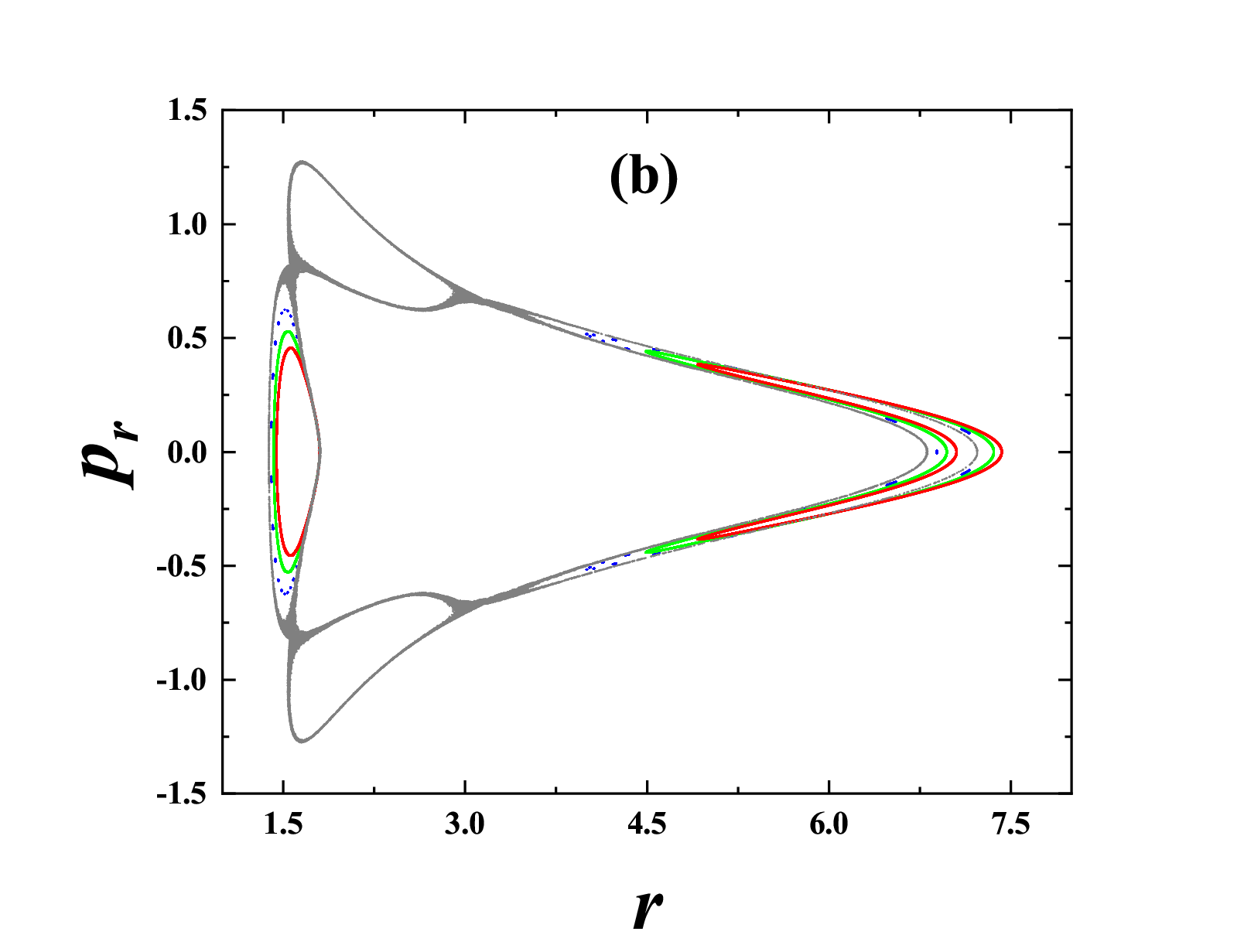}\\[2mm]
    \includegraphics[width=0.49\linewidth]{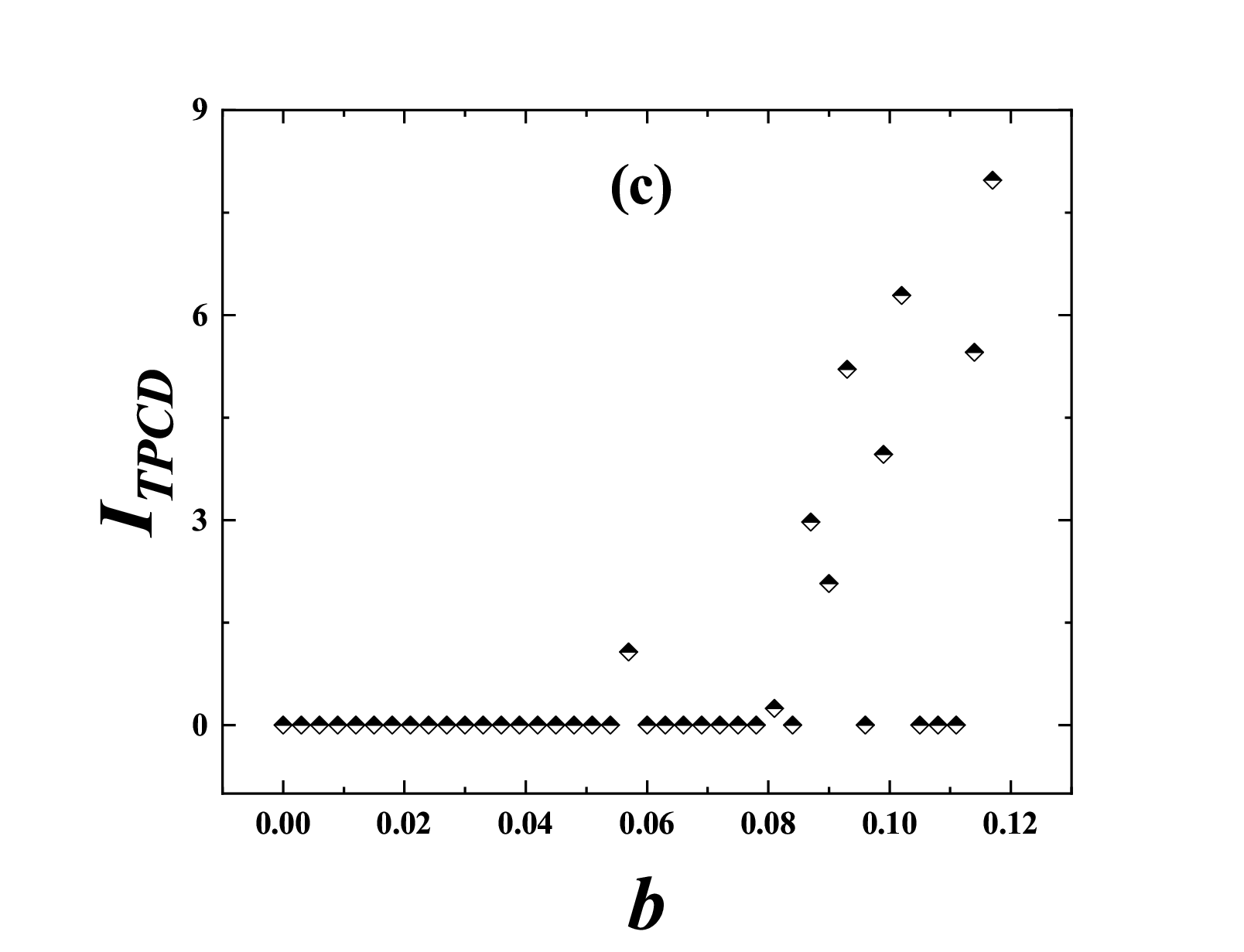}\hfill
    \includegraphics[width=0.49\linewidth]{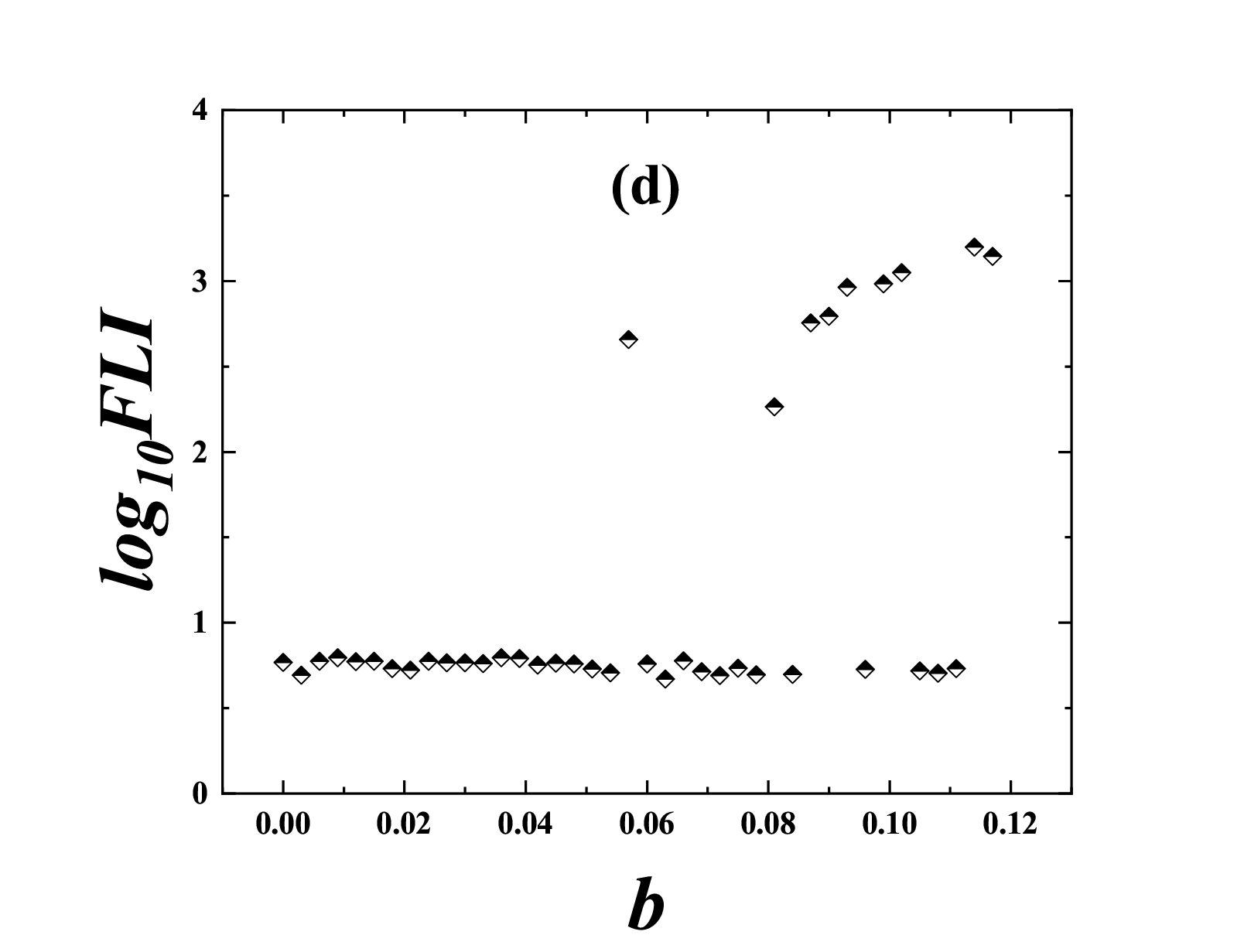}
    \caption{Magnetic-coupling scan and representative Poincar\'e sections for charged-particle motion around a Kerr black hole immersed in an external magnetic field. The fixed parameters are $E=0.905$, $L=2$, $a=0.99$, and $r_0=1.8$, with $\theta_0=\pi/2$, $p_{r0}=0$, and the positive mass-shell solution for $p_{\theta0}$ recalculated at each coupling. In panels (a), (c), and (d), $b=qB/m$ ranges from $0$ to $0.117$ in increments of $0.003$. (a) Radial-to-polar event-count ratio $N_r/N_\theta$, with the radial count in the numerator. The extended plateau at $1/2$ is consistent with a $1{:}2$ radial-to-polar frequency ratio under the event--phase correspondence. (b) Representative Poincar\'e sections in the $(r,p_r)$ plane at $\theta=\pi/2$ with $\dot\theta>0$, showing two principal island regions and nearby scattered points. The gray, blue, green, and red points correspond to $b=0.057$, $0.054$, $0.051$, and $0.048$, respectively. (c) Turning-point count-discrepancy indicator $I_{\mathrm{TPCD}}$. (d) $\log_{10}(\mathrm{FLI})$ for the same magnetic-coupling scan.}
    \label{fig:magnetic_scan_tpcd}
\end{figure}

Figure~\ref{fig:magnetic_scan_tpcd} scans 40 couplings $b=qB/m$ from $0$ to $0.117$ in increments of $0.003$ at fixed $r_0=1.8$. Panel (a) shows a $1/2$ plateau from $b=0.009$ to $0.054$, where TPCD is zero and FLI remains small. The red, green, and blue sections in panel (b), at $b=0.048$, $0.051$, and $0.054$, display two principal island regions consistent with a $1{:}2$ radial-to-polar resonance structure.

Panels (c) and (d) of Fig.~\ref{fig:magnetic_scan_tpcd} show that TPCD and FLI also give consistent results in the magnetic-coupling scan. Both indicators identify nine chaotic trajectories, at $b=0.057$, $0.081$, $0.087$, $0.090$, $0.093$, $0.099$, $0.102$, $0.114$, and $0.117$, while the remaining 31 trajectories are regular. Regular and chaotic intervals alternate as the magnetic coupling increases, rather than exhibiting a monotonic transition to chaos. In particular, the sampled points from $b=0.105$ to $0.111$ form a regular window. The representative Poincar\'e sections in Fig.~\ref{fig:magnetic_scan_tpcd}(b) further support these results: the gray trajectory at $b=0.057$ has a more complicated distribution than the nearby regular trajectories, in agreement with its larger TPCD and FLI.

Because TPCD is a finite-time diagnostic, we additionally extend the integration of the trajectory at $b=0.081$. At $T=10^8$, the value becomes $2.44842$. This result shows that a chaotic trajectory can retain a comparatively stable cycle-count relation over a shorter interval, while longer evolution reveals a substantially larger cumulative discrepancy.

The two parameter scans demonstrate that TPCD effectively distinguishes the regular and chaotic trajectories studied here, with results consistent with FLI at all 80 sampled points and further support from representative Poincar\'e sections. The extended integration also shows that a chaotic trajectory may retain a relatively stable event-count relation over a shorter interval, while a longer observation reveals a substantial cumulative discrepancy. Observation time is therefore an important factor in the diagnostic performance of TPCD.
\section{Schwarzschild--Melvin photon dynamics}\label{sec:melvin}

The Schwarzschild--Melvin solution incorporates the magnetic field into the spacetime geometry  \cite{Ernst:1976,LimaJunior:2021,Li:2018wtz}. We consider photons following null geodesics of the metric
\begin{equation}
ds^2=\mathcal L^2\left[-F\,dt^2+\frac{dr^2}{F}+r^2d\theta^2\right]
+\frac{r^2\sin^2\theta}{\mathcal L^2}\,d\phi^2,
\label{eq:sm_metric}
\end{equation}
where $F(r)=1-2/r$ and $\mathcal L(r,\theta)=1+B^2r^2\sin^2\theta/4$, with the black-hole mass set to unity. Here $B$ is the magnetic parameter of the geometry, and $E=-p_t$ and $L=p_\phi$ are the conserved photon energy and axial angular momentum.

\begin{figure}[p]
\centering
\includegraphics[width=0.49\linewidth]{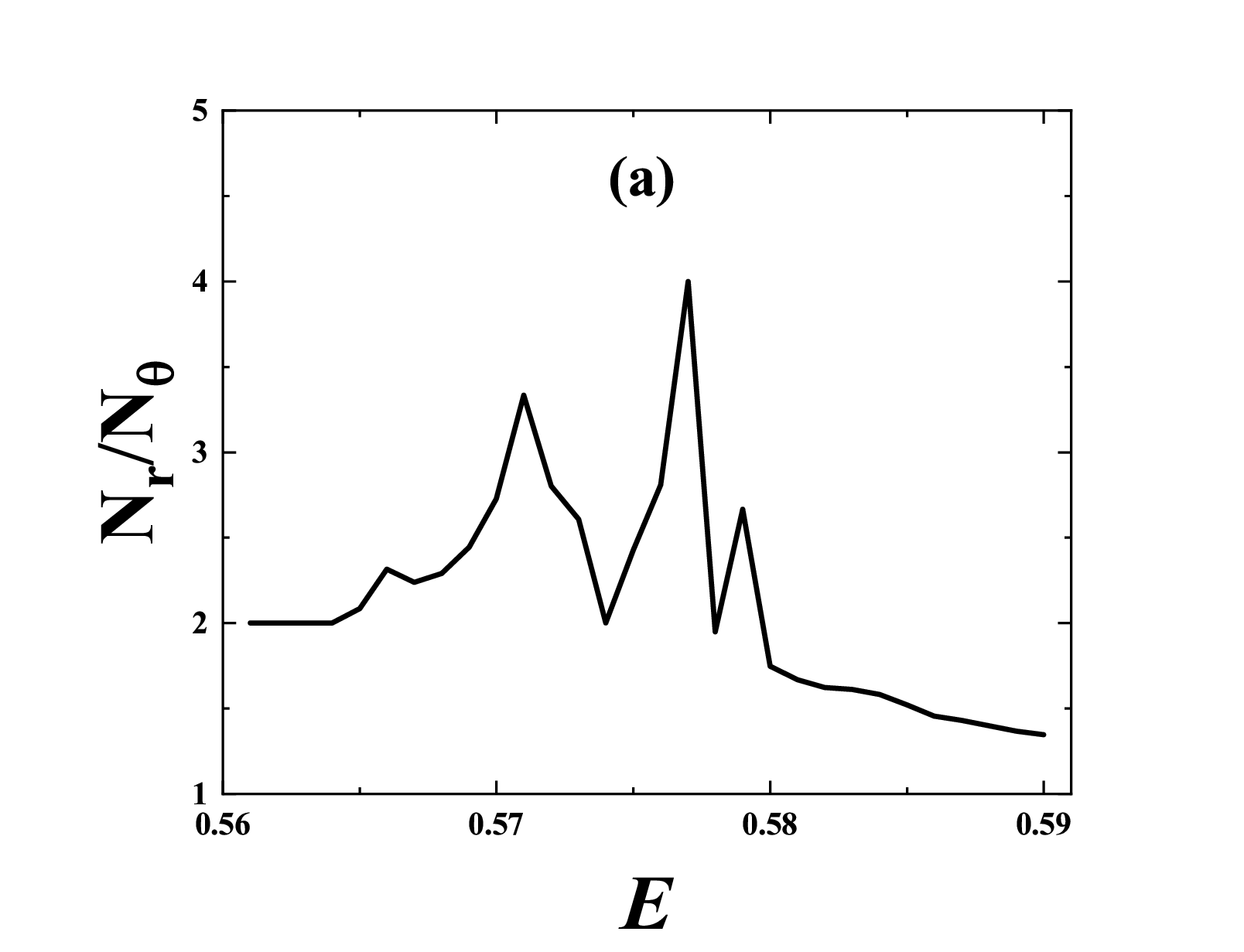}\hfill
\includegraphics[width=0.49\linewidth]{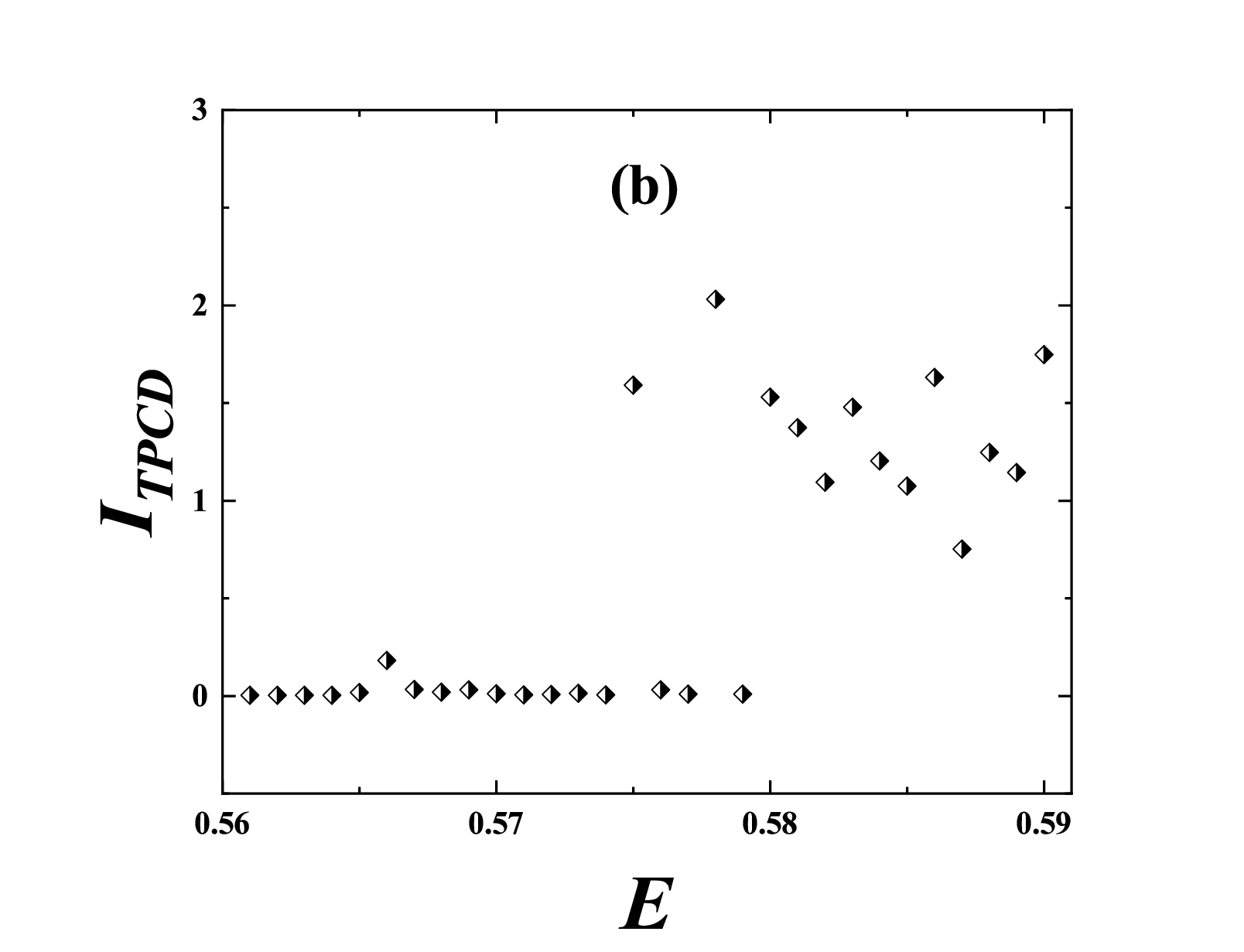}\\[2mm]
\includegraphics[width=0.49\linewidth]{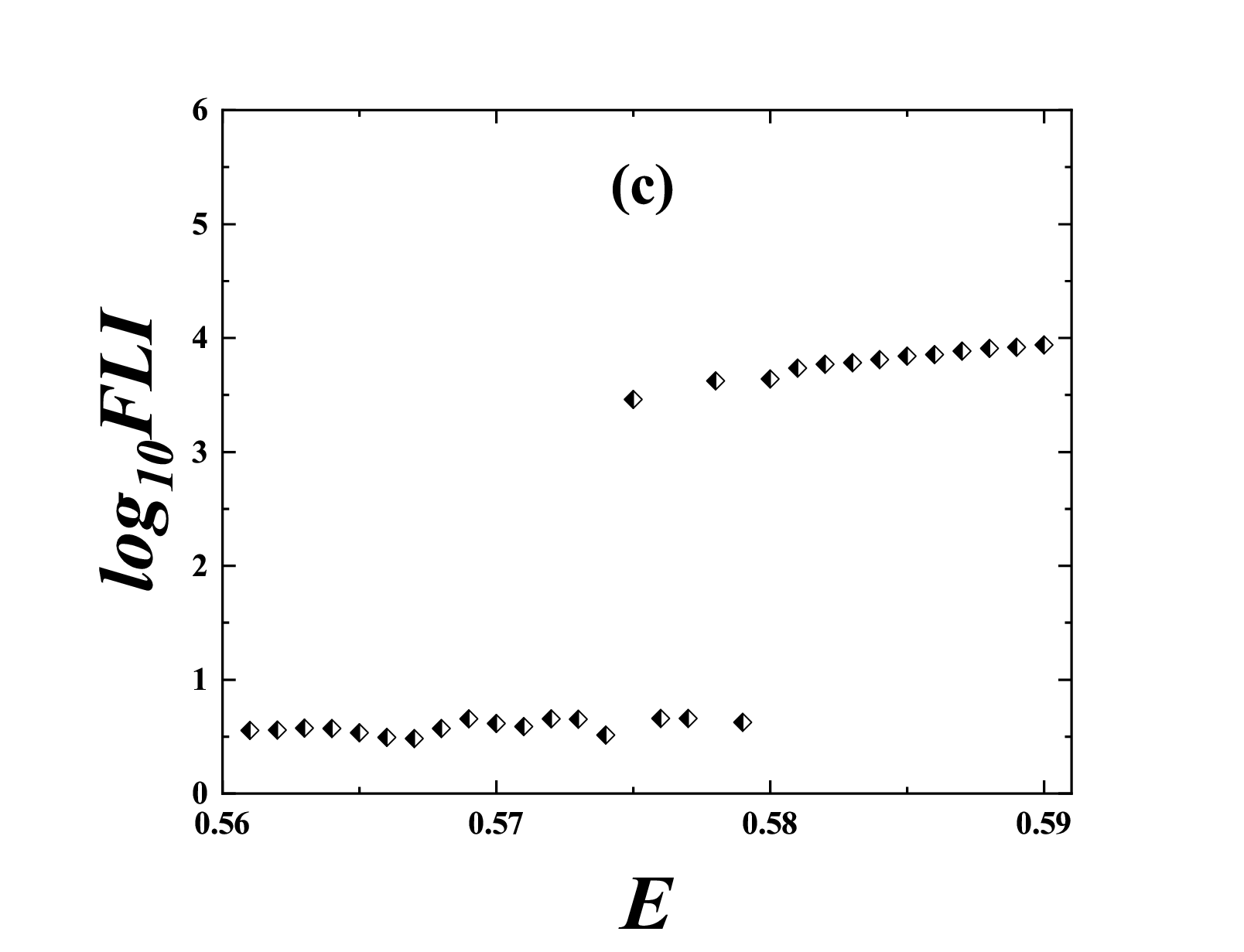}\hfill
\includegraphics[width=0.49\linewidth]{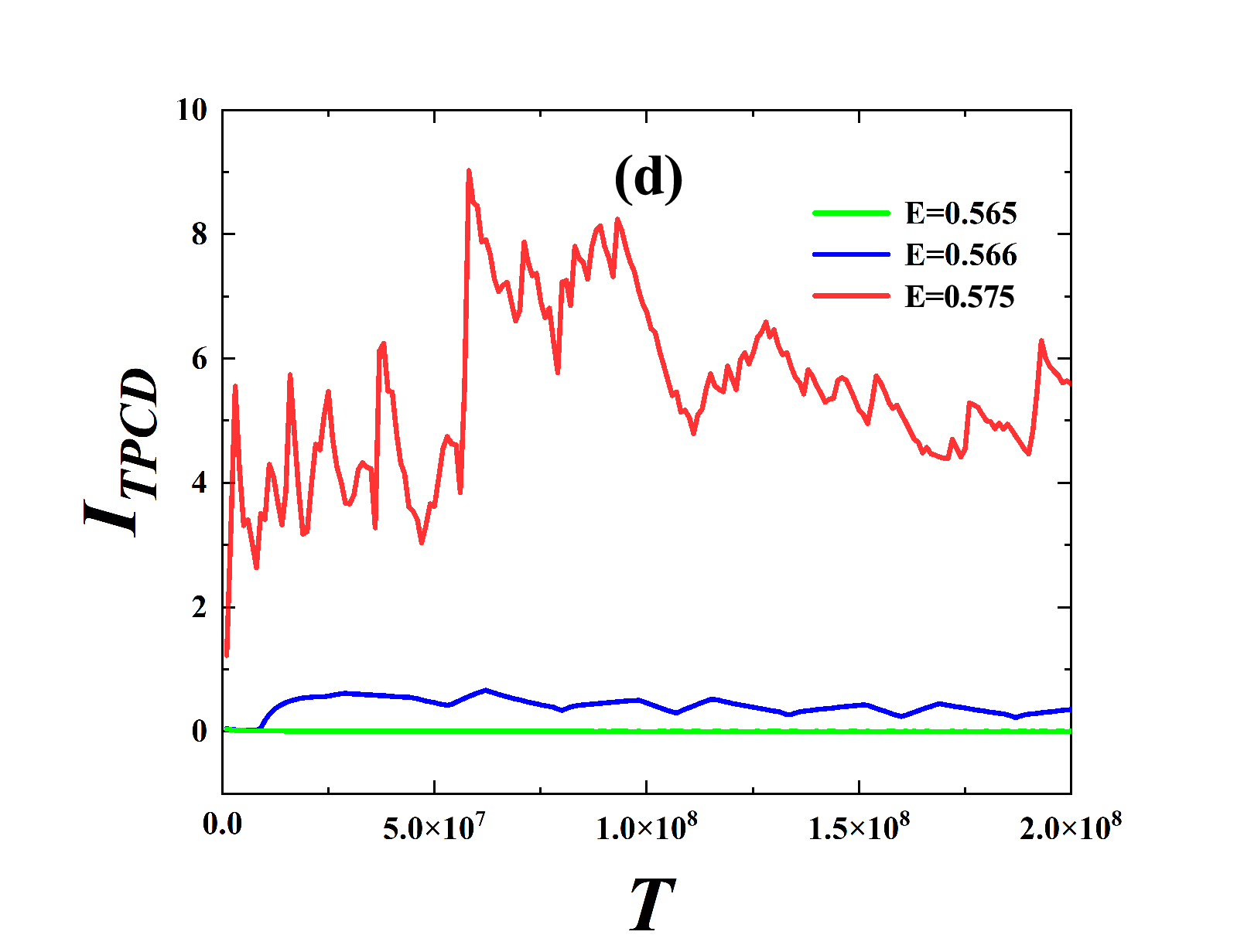}
\caption{Event-count ratios, TPCD, and FLI for Schwarzschild--Melvin photons with $L=4$, $B=0.1$, $r_0=10.656338631529096$, $\theta_0=\pi/2$, $\phi_0=0$, and $p_{r0}=0$. The positive $p_{\theta0}$ is determined separately at each energy by the null constraint. Panels (a--c) scan $E$ from $0.561$ to $0.590$ in steps of $0.001$: (a) radial-to-polar event-count ratio $N_r/N_\theta$; (b) TPCD; (c) $\log_{10}(\mathrm{FLI})$. Panels (a) and (b) use an affine observation interval $T=10^7$, and panel (c) uses $T=10^6$. Panel (d) shows cumulative-record TPCD up to affine interval $T=2\times10^8$ for $E=0.565$ (green), $0.566$ (blue), and $0.575$ (red). Every time sample uses its own mean count over the complete radial cycles available by that time. The ordinate in panel (d) is normalized TPCD, not $R_{\max}$.}
\label{fig:melvin_tpcd}
\end{figure}

Figure~\ref{fig:melvin_tpcd}(a--c) shows an energy scan over 30 values $E=0.561,0.562,\ldots,0.590$, with $L=4$, $B=0.1$, $r_0=10.656338631529096$, $\theta_0=\pi/2$, $\phi_0=0$, and $p_{r0}=0$. At each energy, the positive initial polar momentum is determined by the null constraint $g^{\mu\nu}p_\mu p_\nu=0$. The null geodesic equations are integrated using the same eighth-order Runge--Kutta scheme with affine step size $h=1$; all times $T$ in this section denote elapsed affine parameter in this momentum normalization. The event-count ratio and TPCD in panels (a) and (b) use an observation interval $T=10^7$, whereas the FLI in panel (c) uses $T=10^6$. The radial-to-polar event-count ratio in panel (a) has an approximately constant low-energy segment near 2, followed by pronounced changes and local peaks. As in the Kerr examples, the mean event-count ratio records the relative event-production rate and does not itself classify chaos. The TPCD and FLI panels show broadly corresponding low- and high-indicator regions, including high values at $E=0.575$ and $0.578$ and throughout the displayed segment $E\geq0.580$.

Unlike the clear separation observed in the Kerr scans, the trajectory at $E=0.566$ in the Schwarzschild--Melvin system has a low FLI, but its finite-time TPCD is appreciably larger than those of nearby regular trajectories. This shows that regular trajectories may also exhibit a degree of cumulative count discrepancy over a finite observation interval. According to the theoretical results of Sec.~\ref{sec:regular_tori}, this behavior is compatible with a larger finite discrepancy bound $K$ on a regular torus. It is therefore necessary to examine the long-term evolution of trajectories whose TPCD is elevated but which do not yet exhibit clear signatures of chaos.

Figure~\ref{fig:melvin_tpcd}(d) shows the evolution of TPCD up to $T=2\times10^8$ for three representative trajectories: the ordinary regular trajectory at $E=0.565$, the special regular trajectory at $E=0.566$, and the chaotic trajectory at $E=0.575$. The green curve remains close to zero throughout the observation interval. By comparison, after initially rising to appreciable values, the blue curve exhibits a clear overall downward trend over long times, despite local fluctuations. This behavior is consistent with the theoretical result in Sec.~\ref{sec:regular_tori} that the normalized discrepancy tends to zero for regular trajectories satisfying its assumptions. Together with its low FLI, this behavior identifies the trajectory as regular. In contrast, the red curve remains at a high level throughout the interval and continues to fluctuate markedly, without an observed long-term trend toward zero, consistent with the chaotic nature indicated by its large FLI.

These three photon trajectories show that an elevated finite-time TPCD in the Schwarzschild--Melvin system does not necessarily imply chaos. Longer observations, together with FLI, help distinguish the special regular trajectory from the chaotic trajectory with persistently large count fluctuations.

\section{Conclusions}\label{sec:conclusions}
We have introduced TPCD, an event-count diagnostic for relativistic Hamiltonian motion with two oscillatory degrees of freedom. It measures the maximum cumulative departure of one turning-event count from its mean production rate per reference cycle, using a single trajectory and requiring neither a neighboring orbit nor a partition of phase space. The same record provides a mean event-count ratio, but the ratio and the cumulative discrepancy contain different information, and TPCD captures the accumulated departures that the ratio alone does not retain.

The theoretical core of the method is a bounded discrepancy under explicit event--phase conditions. Rigid phases with an exact event--phase correspondence give the strict bound $|R_m^{(N)}|<1$, which is an exact counting bound rather than a chaos threshold. For a linearizable regular torus whose selected event phases are bounded, degree-one deformations of the torus angles, the bound relaxes to $R_{\max}(N)\leq2K$, with a constant $K$ that is finite but orbit dependent and need not be small. Both results imply that the normalized indicator $I_{\mathrm{TPCD}}$ tends to zero as the number of events grows, so a sustained decay of TPCD is a signature of regular motion. Under a separate statistical diffusion assumption, centered event-count fluctuations instead obey a Brownian-bridge scaling and can sustain a finite statistical scale of TPCD. This mechanism is offered as a conditional expectation for chaotic motion and is not asserted for all chaotic trajectories.

The numerical applications confirm these expectations. Integrable Kerr motion realizes the strict event--phase correspondence and recovers prescribed radial-to-polar frequency ratios from event counts to within $2.6\times10^{-5}$ for six targets, five rational and one irrational, so the construction is not restricted to commensurate motion. For charged particles around a Kerr black hole in an external test magnetic field, the two parameter scans of 40 trajectories each show that TPCD and the fast Lyapunov indicator identify the same 57 regular and 23 chaotic trajectories, with no disagreement at any of the 80 sampled points; the two groups are widely separated, and representative Poincar\'e sections support the classification. Schwarzschild--Melvin photon trajectories demonstrate the applicability of the method to a different dynamical model, in which the magnetic field changes the geometry itself.

These results also reveal the observation-time dependence of TPCD. In the Kerr charged-particle scans, the TPCD values of regular and chaotic trajectories are clearly separated, which demonstrates the effectiveness of the indicator in identifying orbital chaos. The Schwarzschild--Melvin photon model, however, shows that an elevated finite-time TPCD does not necessarily imply chaos. The special trajectory at $E=0.566$ has a finite-time TPCD appreciably larger than those of neighboring regular trajectories, while its FLI remains small; followed to $T=2\times10^{8}$, its TPCD turns over and trends downward, consistent with the regular-torus bound, whereas the chaotic trajectory at $E=0.575$ remains at a high level and continues to fluctuate. For trajectories with intermediate indicator values and an unclear dynamical character, the long-term trend therefore carries information that a single finite-time value does not, and the classification should rest on that trend together with an independent diagnostic. Since the bound $K$ is orbit dependent, a systematic study of how it varies across a phase-space region, and an extension of the event-count construction to systems with more than two oscillatory degrees of freedom, are natural directions for further work.

\section*{Acknowledgments}
This work is supported by the National Natural Science
Foundation of China (NSFC) under Grant nos. 12235019 and 12275106.


\begin{thebibliography}{99}

\bibitem{Carter:1968rr}
B.~Carter,
``Global structure of the Kerr family of gravitational fields,''
Phys. Rev. \textbf{174}, 1559-1571 (1968)
doi:10.1103/PhysRev.174.1559

\bibitem{Wilkins:1972rs}
D.~C.~Wilkins,
``Bound Geodesics in the Kerr Metric,''
Phys. Rev. D \textbf{5}, 814--822 (1972).
doi:10.1103/PhysRevD.5.814.


\bibitem{Ernst:1976}
F.~J.~Ernst,
``Black holes in a magnetic universe,''
J. Math. Phys. \textbf{17}, 54--56 (1976),
doi:10.1063/1.522781.

\bibitem{LimaJunior:2021}
H.~C.~D.~Lima Junior, P.~V.~P.~Cunha, C.~A.~R.~Herdeiro, and L.~C.~B.~Crispino,
``Shadows and lensing of black holes immersed in strong magnetic fields,''
Phys. Rev. D \textbf{104}, 044018 (2021),
doi:10.1103/PhysRevD.104.044018,
[arXiv:2104.09577 [gr-qc]].


\bibitem{Li:2018wtz}
D.~Li and X.~Wu,
``Chaotic motion of neutral and charged particles in a magnetized Ernst-Schwarzschild spacetime,''
Eur. Phys. J. Plus \textbf{134}, no.3, 96 (2019)
doi:10.1140/epjp/i2019-12502-9
[arXiv:1803.02119 [gr-qc]].


\bibitem{Wu:2003pe}
X.~Wu and T.~y.~Huang,
``Computation of Lyapunov exponents in general relativity,''
Phys. Lett. A \textbf{313}, 77-81 (2003)
doi:10.1016/S0375-9601(03)00720-5
[arXiv:gr-qc/0302118 [gr-qc]].

\bibitem{Wu:2006rx}
X.~Wu, T.~Y.~Huang and H.~Zhang,
``Lyapunov indices with two nearby trajectories in a curved spacetime,''
Phys. Rev. D \textbf{74}, 083001 (2006)
doi:10.1103/PhysRevD.74.083001
[arXiv:1006.5251 [gr-qc]].

\bibitem{Ma:2014aha}
D.~Z.~Ma, J.~P.~Wu and J.~Zhang,
``Chaos from the ring string in a Gauss-Bonnet black hole in AdS5 space,''
Phys. Rev. D \textbf{89}, no.8, 086011 (2014)
doi:10.1103/PhysRevD.89.086011
[arXiv:1405.3563 [hep-th]].




\bibitem{Cao:2024ihv}
W.~Cao, X.~Wu and J.~Lyu,
``Electromagnetic field and chaotic charged-particle motion around hairy black holes in Horndeski gravity,''
Eur. Phys. J. C \textbf{84}, no.4, 435 (2024)
doi:10.1140/epjc/s10052-024-12804-8
[arXiv:2404.19225 [gr-qc]].




\bibitem{Liu:2026npn}
B.~H.~Liu, D.~Z.~Ma and Z.~M.~Xu,
``Chaos in Ay{\'o}n-Beato-Garc{\'\i}a black hole coupled with cloud of strings,''
Nucl. Phys. B \textbf{1027}, 117482 (2026)
doi:10.1016/j.nuclphysb.2026.117482



\bibitem{Liu:2026mur}
Z.~Liu and W.~Cao,
``Chaotic dynamics of charged particles near weakly magnetized black holes in Einstein{\textendash}ModMax theory,''
Eur. Phys. J. Plus \textbf{141}, no.4, 417 (2026)
doi:10.1140/epjp/s13360-026-07645-1
[arXiv:2604.21622 [gr-qc]].



\bibitem{Wang:2026whj}
L.~Wang and X.~Wu,
``Discussion on the equivalence of two relativistic point-particle Lagrangians,''
Eur. Phys. J. C \textbf{86}, no.4, 369 (2026)
doi:10.1140/epjc/s10052-026-15594-3
[arXiv:2604.10876 [gr-qc]].





\bibitem{Xu:2026oic}
Z.~Xu, D.~Ma and K.~Li,
``Effects of electromagnetic potential on chaos in Kerr-MOG black holes,''
Eur. Phys. J. C \textbf{86}, no.3, 295 (2026)
doi:10.1140/epjc/s10052-026-15546-x





\bibitem{Lu:2026kcm}
J.~Lu and X.~Wu,
``Third type of spacetime with the coexistence of integrability and non-integrability,''
Eur. Phys. J. C \textbf{86}, no.3, 256 (2026)
doi:10.1140/epjc/s10052-026-15482-w
[arXiv:2603.12674 [gr-qc]].





\bibitem{Li:2026ocp}
K.~Li, D.~Z.~Ma and Z.~M.~Xu,
``Chaotic Motion of Strings in a Quantum-Corrected AdS Reissner{\textendash}Nordstr{\"o}m Black Hole,''
Universe \textbf{12}, no.2, 57 (2026)
doi:10.3390/universe12020057





\bibitem{Zhang:2025qia}
Q.~Zhang and X.~Wu,
``Chaos of Charged Particles in Quadrupole Magnetic Fields Under Schwarzschild Backgrounds,''
Universe \textbf{11}, no.7, 234 (2025)
doi:10.3390/universe11070234
[arXiv:2507.12752 [gr-qc]].





\bibitem{Liu:2025izb}
C.~Liu and X.~Wu,
``Discussions on the integrable dynamics of charged particles around Kerr-Newman black holes in combined gravitational and electromagnetic fields,''
Phys. Lett. B \textbf{867}, 139616 (2025)
doi:10.1016/j.physletb.2025.139616





\bibitem{Xu:2024ble}
Z.~Xu, Z.~M.~Xu, D.~Ma, D.~Z.~Ma, W.~Cao, W.~F.~Cao and K.~Li,
``Chaotic motion of charged test particles in a Kerr-MOG black hole with explicit symplectic algorithms,''
Eur. Phys. J. C \textbf{85}, no.7, 770 (2025)
doi:10.1140/epjc/s10052-025-14425-1
[arXiv:2412.06122 [gr-qc]].





\bibitem{Lu:2024srb}
J.~Lu and X.~Wu,
``Effects of Two Quantum Correction Parameters on Chaotic Dynamics of Particles near Renormalized Group Improved Schwarzschild Black Holes,''
Universe \textbf{10}, no.7, 277 (2024)
doi:10.3390/universe10070277
[arXiv:2406.18943 [gr-qc]].


\bibitem{Das:2024iuf}
S.~Das, S.~Dalui and R.~Samanta,
``Near-horizon chaos beyond Einstein gravity,''
Phys. Rev. D \textbf{110}, no.12, 124037 (2024)
doi:10.1103/PhysRevD.110.124037
[arXiv:2405.09945 [gr-qc]].


\bibitem{Yi:2020shw}
M.~Yi and X.~Wu,
``Dynamics of charged particles around a magnetically deformed Schwarzschild black hole,''
Phys. Scripta \textbf{95}, no.8, 085008 (2020)
doi:10.1088/1402-4896/aba4c2




\bibitem{Chen:2016tmr}
S.~Chen, M.~Wang and J.~Jing,
``Chaotic motion of particles in the accelerating and rotating black holes spacetime,''
JHEP \textbf{09}, 082 (2016)
doi:10.1007/JHEP09(2016)082
[arXiv:1604.02785 [gr-qc]].









\bibitem{Cao:2024bjk}
W.~Cao, Y.~Huang and H.~Zhang,
``Screen chaotic motion by Shannon entropy in curved spacetimes,''
Eur. Phys. J. C \textbf{85}, no.5, 568 (2025)
doi:10.1140/epjc/s10052-025-14310-x
[arXiv:2410.20870 [gr-qc]].

\bibitem{Cao:2024rvo}
W.~Cao, Y.~Huang and H.~Zhang,
``Mutual information for a particle pair and its application to diagnose chaos in curved spacetime,''
Phys. Rev. E \textbf{111}, no.5, 054207 (2025)
doi:10.1103/PhysRevE.111.054207
[arXiv:2412.16931 [gr-qc]].

\bibitem{Cao:2026qvy}
W.~Cao, S.~Chen and H.~Zhang,
``Time-reversed Shannon entropy as a chaos indicator for nonintegrable systems,''
Phys. Rev. D \textbf{113}, no.7, 7 (2026)
doi:10.1103/r9df-spp9
[arXiv:2601.18422 [gr-qc]].




\bibitem{Kopacek:2010yr}
O.~Kopacek, V.~Karas, J.~Kovar and Z.~Stuchlik,
``Transition from Regular to Chaotic Circulation in Magnetized Coronae near Compact Objects,''
Astrophys. J. \textbf{722}, 1240-1259 (2010)
doi:10.1088/0004-637X/722/2/1240
[arXiv:1008.4650 [astro-ph.HE]].








\bibitem{Hinderer:2008dm}
T.~Hinderer and E.~E.~Flanagan,
``Two timescale analysis of extreme mass ratio inspirals in Kerr. I. Orbital Motion,''
Phys. Rev. D \textbf{78}, 064028 (2008)
doi:10.1103/PhysRevD.78.064028
[arXiv:0805.3337 [gr-qc]].

\bibitem{Kerachian:2023oiw}
M.~Kerachian, L.~Polcar, V.~Skoup{\'y}, C.~Efthymiopoulos and G.~Lukes-Gerakopoulos,
``Action-angle formalism for extreme mass ratio inspirals in Kerr spacetime,''
Phys. Rev. D \textbf{108}, no.4, 044004 (2023)
doi:10.1103/PhysRevD.108.044004
[arXiv:2301.08150 [gr-qc]].

\bibitem{Mino:2003yg}
Y.~Mino,
``Perturbative approach to an orbital evolution around a supermassive black hole,''
Phys. Rev. D \textbf{67}, 084027 (2003)
doi:10.1103/PhysRevD.67.084027
[arXiv:gr-qc/0302075 [gr-qc]].

\bibitem{Drasco:2003ky}
S.~Drasco and S.~A.~Hughes,
``Rotating black hole orbit functionals in the frequency domain,''
Phys. Rev. D \textbf{69}, 044015 (2004)
doi:10.1103/PhysRevD.69.044015
[arXiv:astro-ph/0308479 [astro-ph]].





\bibitem{Wald:1974np}
R.~M.~Wald,
``Black hole in a uniform magnetic field,''
Phys. Rev. D \textbf{10}, 1680-1685 (1974)
doi:10.1103/PhysRevD.10.1680




\end{thebibliography}
\end{document}